\documentclass{article}
\usepackage{arxiv}

\usepackage[utf8]{inputenc} % allow utf-8 input
\usepackage[T1]{fontenc}    % use 8-bit T1 fonts

\usepackage{graphicx}
\usepackage{amsmath}
\usepackage{hyperref}
\AtBeginDocument{%
  }

\usepackage{listings}
\usepackage{algorithm}
\usepackage{algorithmicx}
\usepackage{algpseudocode}

\newcommand{\tool}{EnsLLM}

\begin{document}

\title{Enhancing LLM Code Generation with Ensembles: \\ A Similarity-Based Selection Approach}

\author{
 Tarek Mahmud \\
  Texas State University\\
%  Texas, USA \\
  \texttt{tarek\_mahmud@txstate.edu} \\
   \And
 Bin Duan \\
  The University of Queensland \\ 
  %Brisbane, Australia \\
  \texttt{b.duan@uq.edu.au} \\
  \And
 Corina Pasareanu \\
  Carnegie Mellon University \\ 
  %California, USA.\\
  \texttt{pcorina@andrew.cmu.edu} \\
  \And
 Guowei Yang \\
  The University of Queensland \\
  %Brisbane, Australia \\
  \texttt{guowei.yang@uq.edu.au} \\
}

% \renewcommand{\shortauthors}{Mahmud et al.}


\maketitle
\begin{abstract}
Ensemble learning has been widely used in machine learning to improve model robustness, accuracy, and generalization, but has not yet been applied to code generation tasks with large language
models (LLMs). We propose an ensemble approach for LLMs in code generation. Instead of relying on the output of a single model, we generate multiple candidate programs from different LLMs and apply a structured voting mechanism to select the most reliable solution.
For voting, we compute syntactic and semantic similarity using CodeBLEU and behavioral equivalence using CrossHair’s differential behavior analysis. By aggregating these similarity scores, we select the program that best aligns with the consensus among the candidates. We show through experiments that our ensemble approach consistently outperforms standalone  
LLMs on the well-known HumanEval and the more challenging LiveCodeBench datasets, achieving an accuracy of 90.2\% and 50.2\%, respectively, on the two datasets. In comparison, the best-performing LLM (GPT-4o) has an accuracy of 83.5\% and 43.4\%, respectively.
Furthermore, even when restricted to free open source models, our method achieves an accuracy of 80.5\%  and 41.6\%, respectively,  demonstrating the viability of our approach in resource-constrained settings.
\end{abstract}


\section{Introduction}
Large Language Models (LLMs) \cite{brown2020language} have significantly advanced automated code generation, enabling models to generate functionally correct program from natural language prompts. Recent models, such as GPT-4 \cite{gpt4}, CodeLlama \cite{codellama}, and DeepSeekCoder \cite{deepseekcoder}, have demonstrated strong performance on benchmark datasets, making them increasingly popular in software development workflows. These models leverage large-scale training on diverse code repositories, allowing them to generate code for a wide range of programming problems across different programming languages. Their ability to generate, complete, and refactor code has led to widespread adoption in software engineering, competitive programming, and AI-assisted development environments.

However, despite these advancements, LLMs are not infallible and may produce incorrect or suboptimal code, leading to syntactic errors, logic mistakes, or missing edge case handling \cite{chen2021evaluating}. The reliability of LLM-generated code depends heavily on the complexity of the problem, the quality of training data, and the model’s ability to generalize beyond its training distribution. Even state-of-the-art models such as GPT-4 achieve only around 82\% accuracy on functional correctness on standard coding benchmarks such as HumanEval, while open-source models such as CodeLlama achieve much lower accuracy (42-50\%) \cite{bubeck2023sparks, ziemniak2023codellama}. These discrepancies highlight the performance gap between proprietary and open-source models, making it challenging to ensure consistent performance across different LLMs.

A key challenge in LLM-based code generation is the unpredictability of failure modes. Studies have found that LLMs tend to favor syntactically plausible solutions, even when they are functionally incorrect \cite{li2022competition}. Moreover, LLMs struggle with long-range dependencies in code, leading to issues when reasoning about complex data structures or multi-step algorithms. OpenAI Codex and GitHub Copilot evaluations have also revealed that LLMs can introduce security vulnerabilities, such as using deprecated APIs, generating unsafe cryptographic implementations, or failing to handle edge cases properly \cite{pearce2022asleep}. These limitations make post-generation validation essential, reinforcing the need for approaches that can systematically assess and filter correct solutions before deployment in real-world applications.

Ensemble learning has been widely used in machine learning to improve model robustness, accuracy, and generalization \cite{mienye2022survey}. Traditional ensembling techniques, such as bagging, boosting, and stacking, have shown success in various domains, including image classification, natural language processing, and anomaly detection. However, they have not been extensively explored in LLM-based code generation. Prior research suggests that generating multiple solutions, albeit with a single model, and selecting the best candidate can significantly improve performance, as seen in OpenAI’s Codex study, where pass@100 accuracy was far higher than pass@1 \cite{chen2021evaluating}. Inspired by these previous works, we propose \tool, an ensemble-based approach that combines the outputs of multiple LLMs and selects the most reliable solution using a structured voting mechanism. Instead of relying on a single model’s output, \tool\ uses multiple candidate programs and applies a novel similarity-based ranking that aims to evaluate syntactic, semantic, and behavioral correctness.

Specifically, \tool\ integrates CodeBLEU \cite{codebleu}, a popular metric for the evaluation of program produced by LLMs,  and differential analysis 
to assess the reliability of the generated program. CodeBLEU 
is used to measure syntactic and semantic similarity between pairs of candidate programs, such that syntactically and semantically similar and logically aligned solutions receive higher rankings. As CodeBLEU uses only static information about a program, we also investigate a complementary {\em execution-based} differential analysis to detect behavioral inconsistencies between candidate pairs and generate counterexamples where the two candidates produce different outputs. In our work, we define a behavioral similarity metric based on the property-based testing tool CrossHair \cite{crosshair}, which is used for differential analysis (but other differential analyses can also be used \cite{hydiff, lahiri2010differential}). Combining these two metrics, \tool\ computes pair-wise scores for all the candidates, aggregates these scores for each candidate, and selects the candidate with the highest score as the output program of the ensemble.

We evaluate \tool\ on two well-established code generation benchmarks: HumanEval \cite{humaneval} and LiveCodeBench \cite{livecodebench}. Our results show that \tool\ consistently outperforms individual LLMs, achieving 90.2\% accuracy on HumanEval and 50.2\% on LiveCodeBench, surpassing the best standalone model in both datasets. 
Additionally, even when restricted to using only free and open-source LLMs, \tool\ achieves 80.5\% on HumanEval and 41.6\% on LiveCodeBench, proving its viability in resource-constrained environments.

The key contributions of this paper are as follows:
\begin{itemize}
    \item We propose an ensemble-based approach for LLM-based code generation, leveraging syntactic, semantic, and behavioral similarity for improved reliability. 
    \item We propose a novel voting mechanism that integrates CodeBLEU and CrossHair to assess syntactic similarity, semantic alignment, and functional correctness of generated program.
    \item We conduct extensive experiments on HumanEval and LiveCodeBench, demonstrating that \tool\ significantly outperforms standalone LLMs. We analyze failure cases and demonstrate that correct programs reinforce each other, supporting our assumption that independent LLMs are less prone to making identical mistakes.
    \item We show that \tool\ remains effective even when restricted to free and open-source LLMs, making it a viable solution for environments with limited access to proprietary models.
\end{itemize}

\section{Background}

% \subsection{LLM for Code Generation}
% Large Language Models (LLMs) have significantly advanced automated code generation by leveraging deep learning techniques and extensive code repositories. LLMs, such as GPT \cite{brown2020language}, BERT \cite{devlin2019bert}, and T5 \cite{raffel2020exploring}, have been widely adopted for various natural language processing tasks, including machine translation, text summarization, and question-answering. Specialized LLMs for code generation, including OpenAI’s Codex \cite{codex}, Google’s Codey, Meta’s Code Llama \cite{codellama}, and Salesforce’s CodeT5 \cite{wang2021codet5}, have been trained on large-scale datasets containing diverse programming languages, algorithms, and software projects. These models employ transformer-based architectures, which use self-attention mechanisms to capture long-range dependencies in both natural and programming languages. This enables them to understand contextual relationships and generate syntactically correct and functionally meaningful code. 

% Additionally, techniques such as fine-tuning and prompt engineering allow these models to specialize in specific programming tasks, from auto-completing functions to generating entire scripts based on natural language descriptions. Despite their successes, challenges such as hallucinations, incorrect logic, and inefficient code persist. This has led us to the exploration of ensemble methods to combine the strengths of multiple LLMs and improve reliability.

\subsection{Ensemble Learning}
Ensemble learning enhances predictive accuracy by combining multiple models to reduce variance, bias, and overfitting. Key techniques include bagging, boosting, stacking, voting, etc.

Bagging \cite{bagging} (Bootstrap Aggregating) trains multiple models on different dataset subsets sampled with replacement and combines their predictions through averaging (regression) or majority voting (classification). Random Forest, an ensemble of decision trees, exemplifies this approach, reducing variance and improving stability.
Boosting \cite{boosting} trains models sequentially, with each correcting the errors of its predecessor. Algorithms like AdaBoost, GBM, and XGBoost iteratively refine predictions, improving accuracy by focusing on hard-to-classify cases.
Stacking \cite{stacking} combines diverse base models using a meta-learner, which determines the best way to integrate predictions for enhanced generalization. Weighted averaging assigns different contributions to models based on performance, further improving stability and accuracy.

Voting \cite{kittler1998combining} ensembles aggregate predictions from multiple models through majority voting (hard voting) or by averaging predicted probabilities (soft voting). This method is particularly effective when individual models have complementary errors, resulting in a more robust final decision. 

Ensemble methods mitigate individual model limitations, making them effective for classification, regression, and generative tasks. This paper proposes \tool\, a voting-based ensemble of LLMs on the code generation task.

\subsection{CodeBLEU}
CodeBLEU \cite{codebleu} is a specialized metric designed to evaluate the quality of automatically generated code, building upon the traditional BLEU (Bilingual Evaluation Understudy) metric by addressing its shortcomings in the context of programming languages. Standard BLEU, widely used in natural language processing, measures n-gram precision—essentially the overlap of word sequences between a generated text and a reference text—providing a simple yet effective way to assess lexical similarity. However, this approach lacks the depth required to understand the structured nature of code, where syntactic correctness and semantic intent are paramount beyond mere token matching. CodeBLEU overcomes these limitations by integrating a multifaceted evaluation framework that captures both the syntactical, semantical and logical aspects of programming, making it particularly suited for assessing code generation tasks.

CodeBLEU’s evaluation includes four elements. First, n-gram precision captures lexical similarity by comparing token sequences (e.g., keywords, identifiers) between generated and reference program. Second, weighted n-gram matching adjusts this by assigning higher importance to key tokens (e.g., function names over punctuation), refining lexical accuracy. Third, Abstract Syntax Tree (AST) similarity evaluates syntactic structure, comparing hierarchical representations of program constructs like loops or function calls to ensure syntactical consistency. Fourth, data flow similarity analyzes variable transformations and dependencies, verifying logical equivalence even across differing implementations. The final CodeBLEU score is a weighted sum of these components, balancing lexical, syntactic, and semantic fidelity.

This multi-faceted metric enables precise ranking and filtering of program outputs, distinguishing high-quality implementations from flawed ones despite superficial similarities. As background, CodeBLEU’s comprehensive approach lays the groundwork for our method, which builds on its strengths to enhance ensemble-based selection of accurate and functionally reliable solutions.

\subsection{CrossHair}
CrossHair \cite{crosshair} is a Python tool that automatically finds counterexamples to assertions in program by using symbolic execution. It works by analyzing the logic of functions, preconditions, and invariants, exploring different execution paths to detect errors that conventional testing might miss. CrossHair integrates Python’s type hints and assert statements to validate function behavior dynamically, making it a useful tool for debugging and verifying correctness.

The diffbehavior feature in CrossHair allows users to compare different versions of a function by executing them with the same inputs and identifying cases where their outputs diverge. It systematically explores possible inputs and reports counterexamples—inputs that lead to unexpected or incorrect behavior in one version but not the other. By highlighting these discrepancies, diffbehavior helps developers detect unintended changes, regressions, or subtle logic errors. This feature is particularly useful when refactoring or optimizing complex functions, ensuring that modifications do not introduce new bugs while maintaining expected behavior.

\section{Approach}

\begin{figure*}[t!]
    \centering
    \includegraphics[width=.8\linewidth]{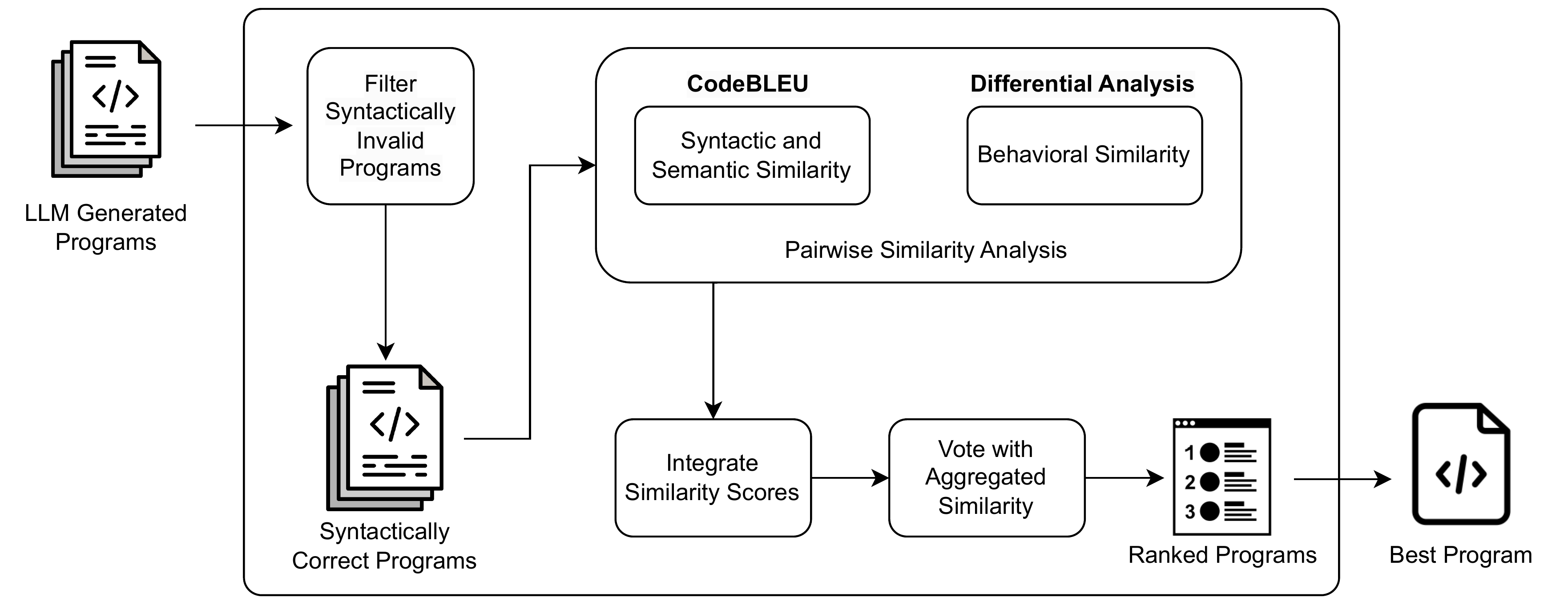}
    \caption{Overview of \tool.}
    \label{fig:overview}
\end{figure*}

%Ensemble learning has been widely used in machine learning to improve model robustness, accuracy, and generalization. Traditional methods such as bagging, boosting, and stacking leverage multiple models to reduce variance and mitigate individual model errors. Inspired by these techniques, 
We propose an ensemble approach for LLM-based code generation that generates multiple candidate programs from different LLMs and applies a voting mechanism to select the most reliable solution.  The main challenge we had to address was: {\em How to do the voting?} We note that although LLMs do not normally compute confidence values, these values can be inferred from the probabilities assigned to each token, during generation. However, as different LLMs are not calibrated, we can not simply use their confidence values for voting. To address this challenge, we propose a voting mechanism that is based on {\em syntactically, semantically, and behaviorally meaningful} pairwise comparison between the programs.

Figure \ref{fig:overview} provides an overview of our approach.
First, it filters out syntactically invalid programs and then selects the most representative solution using similarity-based voting. Since LLM-generated programs often contain hallucinations, incomplete structures, or misplaced tokens, syntactic filtering is a necessary step to prevent misleading similarity calculations and focus the selection process on functionally viable candidates.

To evaluate the similarity between candidate programs, we employ two complementary similarity measures: CodeBLEU for syntactic and semantic similarity and  behavioral similarity measured with differential analysis. CodeBLEU captures lexical, structural, and data-flow relationships between programs, while differential analysis (CrossHair) detects functional inconsistencies by generating counterexamples where two programs produce different outputs. By aggregating these similarity scores, our structured voting mechanism selects the most representative program, as the output of the ensemble.
%one that satisfies both syntactic and semantic correctness and functional reliability.
%, aiming to ensuring that the final output is both syntactically, semantically, and behaviorally sound.

\subsection{Syntactic and Semantic Similarity}
For syntactic and semantic similarity, we use CodeBLEU. As mentioned, CodeBLEU extends traditional BLEU by incorporating program-specific features to evaluate the quality and similarity of generated programs beyond simple token overlap. 
Unlike BLEU, which primarily focuses on n-gram matching in natural language, CodeBLEU introduces additional dimensions tailored for source code: lexical n-gram matching (n\_gram\_weight), syntactic structure matching (syntax\_weight), semantic data-flow analysis (dataflow\_weight), and token weighting (token\_weight). 
%These four similarity matrices help quantify the agreement between different candidate programs, ensuring that syntactically, semantically, and functionally similar programs receive higher similarity scores.
In our work, we only use syntax\_weight  and dataflow\_weight, i.e., assign non-zero weight, as we are primarily concerned with the syntactical and semantic similarity of programs rather than their textual resemblance. By focusing on these two measures, we aim to reduce the influence of superficial lexical similarities while still capturing key syntactical and semantic patterns.

\begin{itemize}
    \item \textbf{Syntactic Structure Matching (syntax\_weight):}
Uses Abstract Syntax Tree (AST) matching to compare the syntactical organization of programs. programs with similar syntax trees are syntactically equivalent, even if their tokens differ. This prevents the penalization of programs with different formatting or minor syntactical variations.
    \item \textbf{Semantic Data-Flow Matching (dataflow\_weight):}
Analyzes data-flow dependencies between variables to assess functional similarity. Two programs are considered equivalent if they manipulate variables and data in the same way, regardless of syntax. This ensures that semantically identical but lexically different implementations receive high similarity scores.
\end{itemize}

CodeBLEU plays a crucial role in our ensemble approach by performing pairwise comparisons between generated programs, ultimately ranking them based on aggregated similarity scores. Given a binary search task, correct implementations—such as the iterative (Listing~\ref{lst:binary_search_iterative}) and recursive (Listing~\ref{lst:binary_search_recursive}) approaches—differ structurally but yield the same results. In contrast, incorrect implementations, such as the linear search (Listing~\ref{lst:linear_search}), introduce inefficiencies or errors that deviate from the intended algorithmic behavior (sometimes small LLMs make this type of mistake).
When performing pairwise similarity checks, CodeBLEU’s syntax-based metric (derived from Abstract Syntax Trees) provides the iterative and recursive binary search implementations a moderate similarity score of 0.6, while its data-flow component captures their logical equivalence with a score of 1.0. These correct implementations reinforce each other in pairwise comparisons, ensuring that they receive consistently high rankings despite their syntactic differences. In contrast, the incorrect implementations receive lower scores when compared to either correct version—for instance, the linear search receives only a syntax similarity of 0.2 with Listing \ref{lst:binary_search_iterative} and 0.3 with Listing \ref{lst:binary_search_recursive} and a data flow similarity of 0.3 with Listing \ref{lst:binary_search_iterative} and 0.2 with Listing \ref{lst:binary_search_recursive}. This low score of the incorrect implementation ensures that correct implementations complement each other syntactically and semantically.

\begin{lstlisting}[language=Python, caption={Iterative Binary Search Implementation}, label={lst:binary_search_iterative}]
def binary_search_iterative(arr, target):
    left, right = 0, len(arr) - 1
    while left <= right:
        mid = (left + right) // 2
        if arr[mid] == target:
            return mid
        elif arr[mid] < target:
            left = mid + 1
        else:
            right = mid - 1
    return -1
\end{lstlisting}

\begin{lstlisting}[language=Python, caption={Recursive Binary Search Implementation}, label={lst:binary_search_recursive}]
def binary_search_recursive(arr, target, left, right):
    if left > right:
        return -1
    mid = (left + right) // 2
    if arr[mid] == target:
        return mid
    elif arr[mid] < target:
        return binary_search_recursive(arr, 
                            target, mid + 1, right)
    else:
        return binary_search_recursive(arr, 
                            target, left, mid - 1)

def binary_search(arr, target):
    return binary_search_recursive(arr, 
                            target, 0, len(arr) - 1)
\end{lstlisting}

\begin{lstlisting}[language=Python, caption={Linear Search Implementation (Incorrect, as the task requires a Binary Search algorithm)}, label={lst:linear_search}]
def linear_search(arr, target):
    for i in range(len(arr)):
        if arr[i] == target:
            return i
    return -1
\end{lstlisting}

A key assumption in our approach is that LLMs do not make identical mistakes when generating an incorrect program. This naturally reduces the similarity between correct and incorrect solutions, such as between the binary search implemntations (Listings~\ref{lst:binary_search_iterative} and \ref{lst:binary_search_recursive}) and the linear search (Listing~\ref{lst:linear_search}), as well as between two incorrect implementations, preventing erroneous programs from being ranked highly. Moreover, similar implementations
will complement each other by reinforcing their shared syntactic and semantic strengths, further boosting their collective score within the ensemble. 

\subsection{Behavioral Similarity}
Syntactical and semantic similarity does not ensure that two programs produce identical runtime behavior. To address this issue, we use execution-based differential analysis. In this work we use CrossHair, a property-based testing tool that detects behavioral inconsistencies by generating counterexamples—inputs that cause two programs to produce different outputs. This allows us to evaluate functional equivalence beyond the static information computed with CodeBLEU.
%just syntactic and lexical similarities.
For each pair of candidate programs, \tool\ runs CrossHair's diffbehavior analysis, systematically exploring edge cases to identify discrepancies. The number of counterexamples serves as an inverse similarity measure:

\begin{itemize}
    \item Zero counterexamples indicate that two programs are behaviorally identical, meaning they produce the same output for all tested inputs.
    \item One or more counterexamples indicate that the programs exhibit different behaviors, highlighting logic errors or functional differences.
\end{itemize}

To quantify behavioral similarity, we introduce a metric that evaluates the consistency of program outputs using differential analysis. The behavioral similarity metric is defined as follows:

\begin{equation}
\label{eq:beh_similarity}
\begin{split}
    \text{BSim}_n(P_i, P_j) &=  \left(1 - \frac{\text{MIN}(n, ({\text{cex}}(P_i, P_j))}{n} \right)
\end{split}
\end{equation}

\noindent where \( \text{cex}(P_i, P_j) \) represents the number of counterexamples detected where the two programs produce different outputs. As the number of counterexamples can be arbitrarily large, we normalize their impact by capping it at \( n \), a small natural number, preventing extreme values from disproportionately influencing the similarity score. This formulation ensures that the similarity decreases as behavioral inconsistencies increase, while the subtraction from 1 prioritizes programs with fewer functional differences.

By incorporating information about behavioral similarity, we aim to ensure that functionally inconsistent programs receive lower similarity scores, improving the reliability of the final program selection.

\begin{lstlisting}[language=Python, caption={Incorrect Recursive Binary Search Implementation}, label={lst:binary_search_recursive_wrong}]
def binary_search_recursive_wrong(arr, target, left, right):
    if left >= right:
        return -1
    mid = left + (right - left) // 2
    if arr[mid] == target:
        return mid
    elif arr[mid] < target:
        return binary_search_recursive_wrong(arr, 
                                    target, mid, right)
    else:
        return binary_search_recursive_wrong(arr, 
                                    target, left, mid)
\end{lstlisting}

We use differential analysis to complement the computed CodeBLEU  similarity. 
Since syntactic and semantic similarity (computed by CodeBLEU) alone do not guarantee identical input-output behavior, differential testing provides an additional layer of validation. When two correct implementations, such as the iterative and recursive binary search functions (Listings~\ref{lst:binary_search_iterative} and \ref{lst:binary_search_recursive}), are compared, they consistently produce identical outputs for all valid inputs on a sorted list, resulting in no behavioral deviations. This confirms their correctness and reinforces their ranking in the selection process.

In contrast, incorrect implementations exhibit behavioral inconsistencies when compared against correct ones. For instance, the flawed recursive binary search (Listing~\ref{lst:binary_search_recursive_wrong}) deviates from the standard binary search logic by improperly updating the search bounds, potentially leading to infinite recursion or missed target values. When the correct iterative (Listing~\ref{lst:binary_search_iterative}) or recursive binary search (Listing~\ref{lst:binary_search_recursive}) is compared against this incorrect version, differential analysis reveals two counterexamples with both Listings \ref{lst:binary_search_iterative} and \ref{lst:binary_search_recursive}. Incorrect programs, when evaluated against correct implementations, consistently produce counterexamples, leading to lower rankings in the selection process. However, CrossHair (differential analysis) alone is not enough; e.g., it would not differentiate between Listings \ref{lst:binary_search_iterative} and \ref{lst:binary_search_recursive} vs. Listing \ref{lst:linear_search}, because they have the same input-output behaviour (but Listing \ref{lst:linear_search} is not a desirable solution since it does not implement a binary search).

\subsection{Integration of Syntactic, Semantic and Behavioral Similarity}

We define our similarity metric by integrating syntactic, semantic, and behavioral equivalence using CodeBLEU and CrossHair-based differential analysis.

\begin{equation}
\label{eq:similarity}
\begin{split}
   \text{similarity}(P_i, P_j) &= \lambda \cdot \text{CodeBLEU}(P_i, P_j) + (1 - \lambda) \cdot \text{BSim}_n(P_i, P_j)
\end{split}
\end{equation}

\noindent Here \( \lambda \) is a weighting factor that balances syntactic and semantic similarity (measured by CodeBLEU) and behavioral similarity (measured by BSim\(_n\)). CodeBLEU provides a normalized similarity score between 0 and 1, evaluating lexical, structural, and data-flow alignment between programs. BSim\(_n\) complements this by assessing functional correctness, ensuring that programs producing similar outputs are ranked higher.

CodeBLEU and CrossHair complement each other by capturing different aspects of correctness, ensuring a more reliable selection process. Although incorrect, Listing~\ref{lst:binary_search_recursive_wrong} appears more structurally similar to the correct implementations when evaluated using CodeBLEU. Since it follows the recursive binary search structure, it receives a relatively high syntax similarity score when compared with Listings~\ref{lst:binary_search_iterative} and \ref{lst:binary_search_recursive}. Additionally, its data flow similarity score remains moderate, as variable transformations in the recursive calls resemble those in the correct implementation, even though they are flawed. This could lead to a higher aggregated CodeBLEU score in pairwise comparisons, potentially misleading the ranking if used in isolation.

However, CrossHair's counterexample analysis helps mitigate this issue by directly verifying behavioral correctness. While Listing~\ref{lst:linear_search} is an entirely different algorithm and thus receives low syntax and data flow similarity scores in CodeBLEU, its simplicity aids CrossHair in validating expected versus actual outputs. Since linear search consistently finds the target, it does not produce any counterexample with Listing \ref{lst:binary_search_iterative} and \ref{lst:binary_search_recursive}, but produces 2 counterexamples in differential analysis with Listing~\ref{lst:binary_search_recursive_wrong}, which fails under certain cases due to incorrect index updates.  This ensures that even among incorrect implementations, CrossHair effectively differentiates functional correctness.
By integrating CodeBLEU’s structural and semantic similarity with CrossHair’s behavioral validation, our approach ensures that one of the correct implementations, Listing~\ref{lst:binary_search_recursive}, is ranked highest. This synergy between CodeBLEU and CrossHair allows us to filter out misleadingly high-scoring but incorrect solutions while ensuring that the most robust implementation is selected.

\subsection{Voting with Aggregated Similarity}

Finally, for each candidate program, we compute an aggregated similarity score by summing its similarity scores with all the other programs in the set. This score quantifies the program’s alignment with the consensus among the generated candidates, ensuring that syntactically, semantically, and behaviorally consistent programs are prioritized. 

\begin{equation}
\text{aggregated\_similarity}(P_i) = \sum_{j \neq i} \text{similarity}(P_i, P_j)
\end{equation}

\begin{algorithm}[t!]
\caption{Voting with Aggregated Pairwise Similarity}
\label{alg:ensemble_selection}
\begin{algorithmic}[1]

\Require List of programs \( P = \{P_1, P_2, ..., P_n\} \)
\Ensure The program \( P^* \) with the highest aggregated similarity

%\State \textbf{//Step 1: Compute Pairwise Similarity}
\For{each \( P_i \) in \( P \)}
    \For{each \( P_j \) in \( P \) where \( i \neq j \)}
        \State \( \text{similarity}[i][j] \gets \text{Compute\_Similarity}(P_i, P_j) \)
    \EndFor
\EndFor

%\State \textbf{//Step 2: Compute Aggregated Similarity}
\For{each \( P_i \) in \( P \)}
    \State \( \text{agg\_sim}[i] \gets 0 \)
    \For{each \( P_j \) in \( P \) where \( i \neq j \)}
        \State \( \text{agg\_sim}[i] \gets \text{agg\_sim}[i] + \text{similarity}[i][j] \)
    \EndFor
\EndFor

%\State \textbf{//Step 3: Identify Best Candidates}
\State \( \text{best\_programs} \gets \{ P_i \mid \text{agg\_sim}[i] = \max(\text{agg\_sim}) \} \)

\State \( P^* \gets tie\_breaking(\text{best\_programs}) \)

\State \Return \( P^* \)

\end{algorithmic}
\end{algorithm}

Algorithm~\ref{alg:ensemble_selection} shows our structured voting-based selection process to identify the most reliable program from a set of candidates. It first computes pairwise similarity between programs and stores the results in a similarity matrix (Lines $1-5$). Then, it calculates each program’s aggregated similarity score by summing its similarity values with all others ((Lines $6-11$). The program with the highest aggregated similarity will be selected as the best program. If multiple programs have the highest score, a tie-breaking function is applied to determine the final selection. In case of a tie in aggregated similarity, the number of CrossHair counterexamples is used as a tiebreaker, prioritizing the program with fewer counterexamples, and if still tied, a random selection ensures a deterministic choice while maintaining diversity in program selection.

Selecting the program with the highest aggregated similarity score ensures that the chosen solution is the most consistent, reliable, and error-resilient among all generated candidates. By prioritizing the program that aligns most closely with the consensus, we reduce the impact of outliers—programs that significantly deviate from common patterns due to syntax errors, logic flaws, or behavioral inconsistencies. Additionally, this approach enhances syntactical, semantical, and behavioral reliability, as the selected program not only shares key syntactic and semantic features with other high-quality candidates but also exhibits minimal behavioral differences, ensuring correct execution. Furthermore, programs that contain incorrect logic or unintended variations are naturally penalized with lower similarity scores, making them less likely to be chosen. As a result, our selection process effectively filters out unreliable outputs while reinforcing the selection of functionally correct and syntactically and semantically sound programs.

\subsection{Limitations}

While our approach significantly improves code generation accuracy, it has several limitations. First, if none of the candidate programs contain a correct solution, our method cannot generate a correct program on its own; it can only select from the available candidates. This makes its effectiveness highly dependent on the quality of the generated outputs from the LLMs. Also, if the assumption, that LLMs do not make identical mistakes, is wrong and multiple models generate similar incorrect outputs, our selection may fail to filter them out.

Second, if after the initial syntactical correctness filtering, only two programs remain, our pairwise similarity-based voting mechanism fails to differentiate between them, as they will always receive identical aggregated scores. This limitation arises because our approach relies on relative ranking rather than absolute correctness evaluation. 

Third, in cases where multiple programs receive similar aggregated similarity scores, our tie-breaking mechanism prioritizes the program with fewer CrossHair counterexamples, but this does not always guarantee correctness. Since CrossHair only identifies behavioral differences based on its generated test cases, it may fail to capture deeper functional issues, leading to ambiguities in the final selection.

\section{Evaluation}

\subsection{Experimental Setup}
To evaluate \tool\, we conduct extensive experiments on two well-established program generation benchmarks: HumanEval and LiveCodeBench. The goal is to assess the effectiveness of our ensemble approach compared to standalone Large Language Models (LLMs) in selecting functionally correct program.

\subsubsection{Research Questions}
%\leavevmode\par
Our evaluation is structured around three key research questions.
\begin{itemize}
    \item
{\textbf{RQ1: How does \tool\ perform compared to standalone LLMs?}} 
    \\ We measure functional correctness using pass@1 metrics and evaluate if \tool\ improves the overall success rate over standalone models.
    \item
{\textbf {RQ2: How do CodeBLEU and CrossHair individually contribute to the accuracy of \tool?}} 
    \\ An ablation study is conducted to assess the independent impact of CodeBLEU (syntactical and semantical similarity) and CrossHair (behavioral equivalence testing) in selecting correct outputs.
    \item
{\textbf{RQ3: How much accuracy can be gained using only free LLMs?}} 
    \\ We analyze whether \tool\ can effectively leverage only open-source models (e.g., OpenChat, CodeLlama, DeepSeekCoder) while maintaining high accuracy.
\end{itemize}

\subsubsection{Benchmarks}
\leavevmode\par
To rigorously evaluate \tool, we utilize two widely recognized benchmarks: HumanEval and LiveCodeBench. These datasets provide diverse programming tasks that assess the ability of large language models (LLMs) to generate functionally correct and syntactically valid program.

\begin{itemize}
    \item \textbf{HumanEval} \cite{humaneval} is a benchmark specifically designed for evaluating the functional correctness of LLM-generated code. It consists of 164 Python programming problems, each presented with a structured format that includes a natural language prompt, a function signature, and a set of hidden test cases used for verification. The problems span a variety of domains, including mathematical computations, string manipulations, data structures, and logical reasoning. Since the primary focus of HumanEval is on correctness rather than efficiency or complexity, the generated program is evaluated using the pass@k metric. This metric determines whether at least one of the top k generated solutions is functionally correct based on the provided test cases. Given its controlled setup and standardized evaluation framework, HumanEval has become a widely adopted benchmark in the automated code generation community.

    \item \textbf{LiveCodeBench} \cite{livecodebench} is a comprehensive benchmark designed to assess the capabilities of LLMs across multiple programming related tasks. The dataset consists of four distinct categories: Code Generation, Self-Repair, Test Output Prediction, and Code Execution. For our evaluation, we focus exclusively on the Code Generation subset, which contains 511 programming problems covering a wide range of coding scenarios. These problems are more diverse than those in HumanEval, as they include real-world coding tasks, API usage, algorithmic challenges, and data structure manipulations. Additionally, LiveCodeBench introduces more complex problem formulations, requiring models to not only generate syntactically correct code but also ensure proper API usage and logical coherence. Like HumanEval, pass@k is used as the primary evaluation metric, providing a robust way to measure functional correctness. The inclusion of this dataset ensures that \tool\ is tested on a broader range of coding challenges that more accurately reflect real-world software development scenarios.
\end{itemize}

\subsubsection{Baseline Models}
\leavevmode\par
We compare \tool\ against a diverse set of 14 LLMs, including both proprietary and open-source models:
\begin{itemize}
    \item Proprietary LLMs: GPT-4o, GPT-4, GPT-3.5, CoPilot, Gemini
    \item Open-Source LLMs: OpenChat, CodeBERT, Llama 3.2, Qwen2, Codestral, Gemma 2, DeepSeekCoder, CodeLlama, DolphinCoder
\end{itemize}

Proprietary models (GPT-4o, GPT-4, GPT-3.5, CoPilot, and Gemini) are accessed via their official APIs, providing state-of-the-art code generation and reasoning capabilities. For open-source models (OpenChat, CodeBERT, Llama 3.2, Qwen2, Codestral, Gemma 2, DeepSeekCoder, CodeLlama, and DolphinCoder), we use the Ollama tool and the ollama-python library for execution. Open-source models offer flexibility, transparency, and offline usability, making them viable alternatives to commercial models. This diverse selection allows for a balanced comparison, evaluating both cutting-edge proprietary models and freely available alternatives in the context of code generation.

\subsubsection{Implementation} 
\leavevmode\par
The implementation of \tool\ follows a structured ensembling process that aggregates outputs from multiple Large Language Models (LLMs) to enhance the reliability and correctness of generated program. The process begins with the collection of candidate solutions from 14 different LLMs, each independently generating responses to a given programming problem. This diversity in generation provides a broad spectrum of possible correct implementations, improving the chances of selecting a high-quality solution. However, before proceeding to selection, all generated programs undergo syntax validation using PyLint, ensuring that syntactically invalid solutions are filtered out, thereby reducing erroneous candidates early in the pipeline.

To determine the most functionally correct program, \tool\ employs a similarity-based selection approach, combining CodeBLEU for syntactic and semantic similarity analysis with CrossHair for behavioral correctness evaluation. CodeBLEU assesses the lexical, syntactic, and semantic alignment between different candidate solutions by analyzing their Abstract Syntax Tree (AST) and data flow. This ensures that solutions syntactically and semantically similar to high-quality implementations are ranked higher. Meanwhile, CrossHair complements this by performing counterexample-based behavioral analysis, identifying discrepancies in function outputs when executed with varying inputs. Solutions that remain consistent under different test cases receive a higher ranking. The final selection is based on an aggregated similarity score that integrates these two evaluation mechanisms, ensuring that the chosen solution aligns with both syntactical and semantical correctness and functional reliability. For this experiment, \tool\ is implemented with the value of $\lambda = .5$ and $n = 10$.

To verify the correctness of the selected solution, \tool\ executes it against unit test cases provided in the benchmark datasets. Each programming problem in HumanEval and LiveCodeBench is associated with a set of predefined test cases that assess whether the generated program produces the expected outputs. In this evaluation, we have used pass@1 as the primary metric, meaning that for each programming problem, we generate only one program solution from an LLM. This metric evaluates the probability that the single generated solution is functionally correct by executing it against the predefined unit test cases in the benchmark datasets. Using pass@1 provides a strict and realistic measure of performance, as it reflects the model's ability to generate a correct solution on the first attempt without relying on multiple outputs. This approach aligns well with real-world coding scenarios, where developers typically seek an immediately functional solution rather than generating multiple alternatives for a single problem.

All experiments are conducted on a high-performance computing system equipped with an NVIDIA GeForce RTX 4060 (16GB VRAM), 32GB RAM, and an Intel Core i7 (12th Gen) processor running Windows 11. The use of GPU acceleration ensures the efficient execution of multiple LLMs, significantly reducing computation time for large-scale benchmarking tasks. This setup allows for scalable and reproducible experimentation, ensuring that \tool’s performance is rigorously evaluated under realistic conditions.

\subsection{Experimental Results}

\subsubsection{RQ1: How does \tool\ perform compared to standalone LLMs?} 

\begin{table}[t!]
    \centering
    \caption{Performance of \tool\ compared to standalone LLMs on HumanEval and LiveCodeBench (pass@1). (F) denotes the free LLMs, and the accuracy provided in the braces is achievable accuracy.}
    \label{tab:elfcg_performance}
    \begin{tabular}{|l|c|c|}
        \hline
        \textbf{LLM} & \textbf{HumanEval (\%)} & \textbf{LiveCodeBench (\%)} \\
        \hline
        GPT-4o         & 83.5  & 43.4  \\
        GPT-4          & 82.3  & 42.2  \\
        GPT-3.5        & 76.8  & 39.1  \\
        CoPilot        & 74.4  & 41.6  \\
        OpenChat (F)      & 71.3  & 37.3  \\
        CodeBert (F)      & 68.9  & 37.2  \\
        Llama 3.2 (F)     & 68.9  & 36.7  \\
        Gemini         & 66.5  & 36.1  \\
        Qwen2 (F)         & 62.1  & 36.4  \\
        Codestral (F)     & 59.2  & 37.2  \\
        Gemma 2  (F)      & 52.1  & 31.3  \\
        DeepSeekCoder (F)  & 50.6  & 25.6  \\
        CodeLlama  (F)     & 42.7  & 23.4  \\
        DolphinCoder (F)   & 34.8  & 22.2  \\
        \hline
        \textbf{\tool\ (All)} & \textbf{90.2 (90.9)} & \textbf{50.2 (53.8)} \\
        \hline
        \textbf{\tool\ (Top 5)} & \textbf{87.2 (90.9)} & \textbf{48.3 (53.8)} \\
        \hline
        \textbf{\tool\ (All Free)} & \textbf{80.5 (83.2)} & \textbf{41.6 (44.1)} \\
        \hline
    \end{tabular}
\end{table}

\leavevmode\par
To evaluate the effectiveness of \tool, we compare its performance against standalone LLMs on the HumanEval and LiveCodeBench datasets using the pass@1 metric. The goal is to determine whether \tool\ improves accuracy over individual models by leveraging an ensemble-based selection strategy.

Table \ref{tab:elfcg_performance} presents the results of individual LLMs as well as \tool's performance on HumanEval and LiveCodeBench in 3 modes. The \tool\ (All) mentioned the results of \tool\ when using all the LLMs considered in this study.
Achievable accuracy, mentioned in the braces with the accuracy of \tool, refers to the maximum possible accuracy that the ensemble-based selection approach can attain, assuming an optimal selection strategy. It is defined as the proportion of problems for which at least one of the individual LLMs in the ensemble produces a correct solution. Since \tool\ does not generate new program but rather selects the best candidate from multiple LLM outputs, its upper bound performance is inherently constrained by the presence of correct solutions among the generated candidates.  
Mathematically, if an ensemble of \( N \) LLMs is used, and for a given benchmark dataset, there are \( M \) total problems and \( C \) is the number of problems where at least one LLM generated a correct solution  the achievable accuracy is calculated by:  

\[
\frac{\text{C}}{M} \times 100
\]

This metric provides a theoretical limit on how well \tool\ can perform. In our evaluation, this upper bound is 90.9\% for HumanEval and 53.8\% for LiveCodeBench, meaning that even with a perfect selection mechanism, the system cannot exceed these accuracy values since no correct solution exists for the remaining problems in the set of LLM-generated outputs.

The best-performing standalone models on HumanEval are GPT-4o (83.5\%), followed closely by GPT-4 (82.3\%) and GPT-3.5 (76.8\%). Similarly, for LiveCodeBench, the highest accuracy among standalone models is GPT-4o (43.4\%), with GPT-4 (42.2\%) and CoPilot (41.6\%) ranking next.
\tool, achieved 90.2\% accuracy on HumanEval, surpassing all standalone models, including GPT-4o, which achieved the highest accuracy among individual LLMs at 83.5\%. This result demonstrates that by leveraging multiple models and selecting the most reliable solution, \tool\ can exceed the performance of any single LLM. Additionally, \tool\ achieved 50.2\% accuracy on LiveCodeBench, significantly outperforming the best standalone model, GPT-4o, which reached 43.4\%. Given the challenging nature of LiveCodeBench problems, this improvement highlights the effectiveness of ensemble-based selection in handling diverse and complex coding tasks.

% To investigate the failed cases of the HumanEval benchmark, where we found that in 121 out of 164 cases, at least two models generated correct answers, and in all such cases, \tool\ successfully selected the correct program. This supports our assumption that in our similarity-based ensemble approach, correct programs complement each other, reinforcing their selection. Among the remaining problems, in 28 cases, at least one model generated the correct program, and \tool\ selected 27 of them correctly. The one case where \tool\ failed was due to three LLMs generating similar mistakes in the program, further supporting our assumption that independent LLMs are less prone to making identical mistakes.

By integrating multiple LLMs and applying a structured ensembling strategy, \tool\ provides substantial improvements over individual models, particularly in LiveCodeBench, where it achieves nearly 7 percentage points higher accuracy than the best-performing standalone model. As shown in Table \ref{tab:elfcg_performance}, \tool\ consistently outperforms all standalone LLMs, demonstrating that an ensemble-based selection approach can significantly enhance functional correctness in code generation. The results further emphasize that while standalone LLMs are constrained by their individual capabilities, \tool\ effectively combines their strengths, leading to higher accuracy and more reliable code generation across diverse problem domains.

%{\bf Corina: moved the following 2 paragraphs here}

To further analyze the impact of model selection, we evaluated an alternative configuration, \tool\ (Top 5), which selects the best-performing subset of five models rather than using all available LLMs. This setup achieves an accuracy of 87.2\% on HumanEval and 48.3\% on LiveCodeBench, compared to the achievable accuracy of 90.9\% and 53.8\%, respectively. The results show that even with a limited subset of models, EnsLLM (Top 5) is able to approach the upper bound of achievable accuracy; on the other hand, the results indicate that the less performant models -- not present in the \tool\ (Top 5) but present in \tool\ (All) -- do have a significant contribution in the performance of the ensemble.
%suggesting that optimal model selection plays a crucial role in maximizing performance. Additionally, the smaller gap between EnsLLM (Top 5) and the achievable accuracy in HumanEval compared to LiveCodeBench highlights that more complex coding problems benefit from a broader selection of models.

\begin{table}[t!]
    \centering
    \caption{Ablation study showing the contribution of CodeBLEU and CrossHair to accuracy.}
    \label{tab:ablation_codebleu_crosshair}
    \begin{tabular}{|l|c|c|}
        \hline
        \textbf{$\lambda$} & \textbf{HumanEval (\%)} & \textbf{LiveCodeBench (\%)} \\
        \hline
        1 & 85.9 & 49.5 \\
        .75 & 87.2 & 49.5 \\
        .5 & \textbf{90.2} & \textbf{50.2} \\
        .25 & 89.6 & 48.4 \\
        0 & 89.6 & 47.9 \\
        
        \hline
    \end{tabular}
\end{table}

\begin{table*}[t!]
    \centering
    \caption{Impact of different CodeBLEU weighting configurations on accuracy.}
    \label{tab:codebleu_variants}
    \scalebox{0.8}{
    \begin{tabular}{|l|c|c|}
        \hline
        \textbf{Settings} & \textbf{HumanEval (\%)} & \textbf{LiveCodeBench (\%)} \\
        \hline
        syntax\_weight = 0.25, dataflow\_weight = .25, n\_gram\_weight = 0.25, token\_weight = 0.25 & 85.4 & 46.1 \\
        syntax\_weight = 0.5, dataflow\_weight = .25, n\_gram\_weight = 0.25 & 86.0 & 46.4 \\
        syntax\_weight = 0.25, dataflow\_weight = .5, n\_gram\_weight = 0.25 & 86.6 & 46.4 \\
        syntax\_weight = 0.25, dataflow\_weight = 0.75 & 89.0 & 49.5 \\
        syntax\_weight = 0.5, dataflow\_weight = 0.5 & 89.0 & 49.5 \\
        \hline
    \end{tabular}}
\end{table*}

\subsubsection{RQ2: How do CodeBLEU and CrossHair individually contribute to the accuracy of \tool?} 

\leavevmode\par
Equation \ref{eq:similarity} defines the unified similarity metric, which integrates CodeBLEU and behavioral similarity (BSim\(_n\)) using a weighting factor \(\lambda\) to balance syntactic/semantic similarity and behavioral equivalence. CodeBLEU captures lexical, syntactical, and semantic alignment, while BSim\(_n\) evaluates functional correctness by measuring behavioral differences between candidate programs.

To determine the optimal balance between CodeBLEU and BSim\(_n\), we conducted an ablation study by varying \(\lambda\) and evaluating accuracy on HumanEval and LiveCodeBench (Table \ref{tab:ablation_codebleu_crosshair}). The results show that \(\lambda = 0.5\) achieves the highest accuracy on both datasets, with 90.2\% on HumanEval and 50.2\% on LiveCodeBench. This suggests that giving equal weight to syntactic/semantic similarity and behavioral similarity yields the most robust and reliable selection of generated programs.

When \(\lambda = 1\) (fully relying on CodeBLEU), accuracy drops to 85.9\% on HumanEval and 49.5\% on LiveCodeBench, indicating that ignoring behavioral correctness leads to functional errors despite structural similarity. Conversely, when \(\lambda = 0\) (fully relying on BSim\(_n\)), accuracy reaches 89.6\% on HumanEval but declines to 47.9\% on LiveCodeBench, showing that behavioral checks alone are insufficient without syntactic/semantic alignment. These results confirm that both structural similarity and behavioral correctness are essential, and an equal balance between the two provides the best overall performance.

To gain deeper insights into the impact of CodeBLEU, we explored various weighting strategies for its four similarity components: lexical similarity, syntactic similarity, data-flow similarity, and token importance weighting. We systematically tested all possible combinations of these weights to determine their influence on accuracy. Table \ref{tab:codebleu_variants} presents only the configurations that yielded better results compared to the baseline, where all four components were equally weighted at 0.25 for both datasets. When all four CodeBLEU matrices were equally weighted at 0.25, the accuracy was 85.4\% on HumanEval and 46.1\% on LiveCodeBench. This indicates that while distributing the weights evenly provides reasonable accuracy, it may not be optimal for all problem types.

By adjusting the weights to favor syntactic and data-flow similarity, where syntax\_weight = 0.5 and dataflow\_weight = 0.5, accuracy improved to 89.0\% on HumanEval and 49.5\% on LiveCodeBench. This suggests that prioritizing syntactic alignment and semantic equivalence results in better selection performance, as these aspects more accurately capture correct implementations. Similarly, using syntax\_weight = 0.25 and dataflow\_weight = 0.75 produced similar results, reinforcing that placing greater emphasis on semantic correctness improves overall accuracy.

These results demonstrate that lexical similarity and token importance weighting contribute less to correctness verification, whereas syntax and data-flow analysis are critical for selecting high-quality code. Reducing the influence of lexical similarity prevents the model from favoring textually similar but functionally incorrect solutions, improving the overall robustness of \tool.

The answer to this research question confirms that CrossHair is highly effective for ensuring behavioral correctness, while CodeBLEU is crucial for ranking semantically similar and syntactically well-formed programs. However, using both together yields the best performance, as each metric complements the other. Additionally, optimizing CodeBLEU’s weight distribution—by prioritizing syntax and data-flow similarity over lexical similarity—further improves accuracy.

To determine the optimal value of \( n \), we conducted experiments by varying \( n \) from 5 to 10 and analyzing its impact on accuracy. The results showed that setting \( n \) between 6 and 10 yielded the highest accuracy, while choosing \( n \) below 6 led to a noticeable decline in performance. Specifically, for \( n < 6 \), the accuracy dropped to 89.2\%, indicating that a lower \( n \) fails to capture sufficient behavioral differences, reducing the effectiveness of the selection process.

These findings reinforce that a combined approach leveraging both syntactic and behavioral correctness is essential for achieving the highest accuracy in \tool. By fine-tuning CodeBLEU weights and integrating CrossHair for functional validation, \tool\ effectively enhances code generation reliability across diverse problem domains.

\subsubsection{RQ3: How much accuracy can be gained using only free LLMs?}

\leavevmode\par
To evaluate the effectiveness of \tool\ when restricted to using only free LLMs, we conducted experiments using an ensemble of open-source models rather than proprietary ones like GPT-4o and CoPilot. The goal was to determine whether \tool\ can still achieve competitive performance without relying on high-resource commercial models.

The results, presented in Table \ref{tab:elfcg_performance}, show that when using only free models, the highest-performing standalone model is OpenChat, which achieves an accuracy of 71.3\% on HumanEval and 37.3\% on LiveCodeBench. The achievable accuracy, which represents the upper bound where the best free model is selected for each problem, is 83.2\% for HumanEval and 44.1\% for LiveCodeBench. This indicates that while free models individually underperform compared to proprietary models, an optimal selection strategy can still yield reasonably high accuracy.

\tool\ achieves an accuracy of 80.5\% on HumanEval and 41.6\% on LiveCodeBench using only free models. This demonstrates that even without access to high-end proprietary models, \tool\ can still significantly improve accuracy by leveraging an ensemble of diverse open-source LLMs. Compared to the best individual free model, \tool\ gains nearly 9 percentage points on HumanEval and 4.3 percentage points on LiveCodeBench, highlighting the effectiveness of our ensembling approach in improving functional correctness.

The results confirm that \tool\ can still achieve strong performance using only free models, significantly outperforming individual open-source LLMs. While it does not reach the accuracy of proprietary model ensembles, it remains a viable alternative when commercial models are unavailable. The ensemble selection process is key to improving accuracy, as it compensates for the weaker performance of individual free models by strategically selecting the most functionally correct solutions. This demonstrates that ensemble-based code generation approaches can enhance reliability even in resource-constrained settings, making them practical for scenarios where access to proprietary models is limited.

However, \tool's accuracy in this setting remains lower than the full ensemble (which includes proprietary models) due to the limited capabilities of free models, particularly in handling complex coding tasks in LiveCodeBench. The gap between \tool's performance (80.5\%) and the achievable accuracy (83.2\%) in HumanEval suggests that further optimizations in model selection and weighting strategies could narrow this difference.

\section{Discussions}
\subsection{Qualitative Analysis}
To better understand the behavior of \tool\, we conducted a qualitative analysis of the failed cases on the HumanEval benchmark. Out of 164 problems, we found that in 131 cases, at least two individual models independently generated correct solutions. In all of these instances, \tool\ successfully selected the correct program. This strongly supports our hypothesis that correct outputs reinforce one another in our similarity-based ensemble, making them more likely to be chosen.

For 18 problems (out of the rest 33 problems), only a single model produced the correct solution for each of those problems. Even in these more challenging scenarios, \tool\ correctly identified the correct program in 17 cases, as no incorrect programs made similar mistakes that would have reinforced one another. The one failure occurred when three models (Llama 3.2, Gemma 2, and CodeLlama) produced incorrect programs with similar mistakes, leading \tool\ to select the wrong output. This illustrates a key limitation of the approach: when multiple models converge on similar errors, the ensemble may amplify rather than avoid them. However, such failures were rare and further support the assumption that diverse model behaviors reduce correlated mistakes. For the remaining 15 cases, no correct solutions were available for \tool\ to select.

\subsection{Scalability Analysis}
Collecting outputs sequentially from each of the LLMs takes 94.22 seconds per problem on average, with an associated cost of approximately 0.005 USD per proprietary model. While parallel execution can significantly reduce latency, running all 14 models, especially at scale, poses practical challenges due to resource demands and infrastructure constraints. The ensemble evaluation stage adds further overhead: computing all 91 pairwise similarity scores using CodeBLEU takes 86.21 seconds per problem, and running behavioral checks with CrossHair requires 212.38 seconds. These components are critical to \tool’s accuracy but introduce notable computational cost. In future work, we plan to explore strategies for more efficient ensemble execution, including selective model invocation, partial ensembling, or model pruning. We also aim to investigate system-level solutions for running multiple LLMs concurrently with minimal resource overhead, to make the approach more practical for real-world deployment.

\begin{table}[t!]
    \centering
    \caption{Performance of \tool\ compared to standalone LLMs on HumanEval-Java(pass@1). The accuracy provided in the braces is achievable accuracy.}
    \label{tab:java_performance}
    \begin{tabular}{|l|c|c|}
        \hline
        \textbf{LLM} & \textbf{Accuracy (\%)} \\
        \hline
        GPT-4o         & 90 \\
        GPT-4          & 84  \\
        GPT-3.5        & 78  \\
        CoPilot        & 86  \\
        OpenChat (F)      & 70  \\
        % CodeBert (F)      & z  \\
        Llama 3.2 (F)     & 72  \\
        Gemini         & 86  \\
        Qwen2 (F)         & 72  \\
        Codestral (F)     & 68  \\
        Gemma 2  (F)      & 62  \\
        DeepSeekCoder (F)  & 60  \\
        CodeLlama  (F)     & 64  \\
        DolphinCoder (F)   & 48  \\
        \hline
        \textbf{\tool\ (All)} & \textbf{90 (90)}  \\
        \hline
        % \textbf{\tool\ (Top 5)} & \textbf{87.2 (90.9)} & \textbf{48.3 (53.8)} \\
        % \hline
        % \textbf{\tool\ (All Free)} & \textbf{80.5 (83.2)} & \textbf{41.6 (44.1)} \\
        % \hline
    \end{tabular}
\end{table}

\subsection{Case Study on HumanEval-Java}
To evaluate if \tool\ works for other languages, we conducted a case study on the HumanEval-Java \cite{humaneval-java} benchmark, which closely follows the structure and semantics of HumanEval but is written in Java. For syntactic and semantic similarity, we reused CodeBLEU, which supports multi-language analysis without modification. However, since CrossHair is Python-specific, we choose an alternative approach to approximate behavioral similarity for Java programs.

We employed Kex \cite{kex2025github}, a symbolic execution-based test generation framework for Java, as a substitute for CrossHair in the Java setting. Kex performs path-sensitive analysis by symbolically executing Java bytecode to explore multiple execution paths and automatically synthesize input values that trigger them. For each candidate program $P_i$, we used Kex to generate a suite of test cases that exercise its behavior over a wide range of symbolic inputs. These tests aim to reflect the program's logic and edge cases without requiring manually written specifications.

To compute behavioral similarity between two programs $P_i$ and $P_j$, we leveraged these generated test cases in a cross-execution strategy. Specifically, for every program pair $(P_i, P_j)$, we executed the test suite %originally 
synthesized for $P_i$ on the implementation of $P_j$. We then counted the number of test cases that failed when run against $P_j$—i.e., those that produced exceptions, assertion failures, or incorrect return values. This count was denoted as $cex(P_i, P_j)$ and treated as a proxy for behavioral divergence from $P_i$'s semantics.

To convert this into a normalized similarity measure, we applied the same formulation used in Equation~(1), capping the counterexamples at $n = 6$ to avoid extreme penalty due to overly sensitive or noisy test cases.

Due to the significant effort required to integrate this new workflow, we report here on a preliminary evaluation on the first 50 problems from the HumanEval-Java dataset. We exclude CodeBERT from this experiment as its performance is hindered by the prompt structure used in HumanEval-Java. Unlike benchmarks like HumanEval or LiveCodeBench, HumanEval-Java lacks well-researched prompt engineering, leading CodeBERT, which is a masked language model, to return the input without generating meaningful method bodies.

As shown in Table \ref{tab:java_performance}, EnsLLM achieved an accuracy of 90\%, which matches the achievable accuracy. Although the accuracy of the best-performing standalone model in our experiment, GPT-4o, is also 90\%, EnsLLM selected the correct solution in every case where a correct solution was available. These findings suggest that the core principles of our ensemble method generalize well across programming languages, provided that compatible analysis tools are available or can be adapted.

\subsection{Threats to Validity}
The main threat to internal validity is CrossHair’s effectiveness depends on the complexity of generated programs; if a program involves external dependencies or non-deterministic behavior, CrossHair may fail to detect functional discrepancies. To mitigate these risks, we integrate both CodeBLEU and CrossHair, ensuring that correct implementations reinforce each other while incorrect ones are penalized.

For external validity, our evaluation is limited to HumanEval and LiveCodeBench, which may not fully represent complex, real-world programming challenges. However, these two benchmarks are widely recognized in code generation research and have been used in several similar studies to evaluate the functional correctness of LLM-generated programs. Their extensive use ensures a standardized comparison with existing approaches, mitigating the threat to generalizability. A potential threat to external validity lies in the selection of the 14 LLMs used in our ensemble. While we included a mix of open-source and proprietary models with varied sizes and tuning, the selection was not based on a formal strategy. This may limit the generalizability of our results, particularly regarding robustness to correlated errors. Future work will explore principled selection methods based on model diversity and behavioral variance to strengthen ensemble design.
Another external validity concern is the potential presence of the solution of the datasets in the pretraining sets of some LLMs. While we cannot verify the training data of all models, any such overlap could inflate performance. Although our focus is on the ensemble mechanism rather than generation quality, this remains a limitation.

A potential threat to validity in our study is bias in language model selection, as the set of 14 LLMs we evaluate may not fully represent the entire spectrum of code generation models. While we include a diverse mix of proprietary and open-source models, newer or more specialized models may perform differently, potentially affecting our findings. To mitigate this, we select widely used models, ensuring our comparison aligns with existing research and industry practices. However, future work will expand this evaluation to a broader set of models to further validate generalizability.

\section{Related Work}
Large Language Models (LLMs) have significantly advanced automated code generation, enabling models to generate functionally correct code from natural language prompts. Early research introduced OpenAI Codex, a GPT-based model fine-tuned for code generation, which demonstrated strong capabilities in solving programming problems \cite{chen2021evaluating}. Codex was evaluated on the HumanEval benchmark, achieving a 28.8\% success rate on pass@1 and 70.2\% on pass@100, highlighting the benefits of generating multiple outputs and selecting the best one. This approach led to the development of tools like GitHub Copilot, which integrates LLM-based code generation into modern IDEs \cite{pearce2022asleep}.

Despite these advances, LLMs still exhibit major limitations in correctness, reliability, and efficiency. Studies have shown that even state-of-the-art models like GPT-4 achieve only ~82\% correctness on HumanEval, while open-source models such as CodeLlama perform significantly worse, with accuracy ranging between 42-50\% \cite{bubeck2023sparks, ziemniak2023codellama}. LLMs often generate program that appears syntactically correct but contains logical errors, missing imports, or faulty API usage. Another challenge is security—a study on GitHub Copilot found that 40\% of AI-generated program contained vulnerabilities, raising concerns about blindly adopting LLM outputs in production codebases \cite{pearce2022asleep}.

To mitigate these issues, research has explored strategies such as reinforcement learning from human feedback (RLHF), filtering generated program based on static analysis tools, and incorporating behavioral validation techniques to assess correctness \cite{fan2023large}. In competitive programming, DeepMind’s AlphaCode demonstrated that LLMs can tackle algorithmically complex problems by generating thousands of solutions and filtering them based on test execution results. AlphaCode was able to achieve human-level performance in Codeforces competitions, ranking among the top 54\% of human participants \cite{li2022competition}. This suggests that while LLMs still require validation and post-processing, their ability to generate correct program improves significantly when paired with structured filtering and selection methods.
Our work differs by introducing an ensemble-based approach that selects the most reliable program from multiple LLM-generated outputs using CodeBLEU for syntactic and semantic similarity and CrossHair for behavioral validation.

Beyond code generation, LLMs have demonstrated impressive performance across various natural language processing and coding-related tasks, including code completion \cite{deng2022fuzzing, huang2022prompt, jain2022jigsaw, li2022competition, xu2022systematic}, code synthesis \cite{liu2023your}, and program repair \cite{xia2023automated}. Codex \cite{codex}, the model behind GitHub Copilot \cite{copilot}, has shown promise in automatically translating comments into functional code, significantly improving developer productivity, though minor issues persist. Other LLMs, such as BERT \cite{devlin2018bert, tian2023best}, GPT \cite{gpt4, chatGPT, xia2023keep}, and various domain-specific models \cite{le2023invalidator, paul2023automated}, have also proven their ability to generate syntactically correct and contextually relevant program.

LLMs have also been explored in broader software engineering tasks, including automated bug detection, test case generation, code summarization, and security vulnerability analysis \cite{rahul2023llm, hou2024large}. For instance, they can assist in writing unit tests by generating test cases from function signatures, improving test coverage while reducing developer effort \cite{mueller2023automated}. Additionally, LLMs are being integrated into code refactoring tools, where they suggest optimizations and enhance code readability \cite{fan2023large}. Beyond these applications, LLMs are transforming software engineering workflows, influencing how software is developed \cite{imai2022github, liu2023fill}, enhancing developer productivity \cite{dakhel2023github, ziegler2022productivity}, and assisting in security analysis \cite{pearce2023examining}. However, challenges persist in ensuring correctness, maintainability, and security, requiring further refinement and validation before LLM-generated code can be reliably integrated into real-world software systems.

Test-based selection has been explored to improve code generation. CodeT~\cite{chen2022codet} uses LLM-generated tests to rank candidate programs from the same model based on execution results and agreement. Fakhoury et al.~\cite{fakhoury2024llm} similarly apply unit test feedback in an interactive loop to guide code refinement. While conceptually aligned with our goal of functional selection, these methods rely on a single model, limiting candidate diversity. On the other hand, \tool\ leverages multiple LLMs, reducing correlated errors and enabling more robust selection.

The mixture of experts (MoE) framework \cite{jordan1994hierarchical} combines multiple specialized models with a gating network to select the best expert for each input, using Expectation-Maximization for training. Fedus et al.~\cite{fedus2021switch} scaled MoE to trillion-parameter Switch Transformers with sparse activation for efficient language processing. 
Despite these advancements, computational efficiency remains a challenge when using multiple models for selection and refinement. Also, \tool\ is the first approach that ensembles multiple LLMs for code generation.

\section{Conclusion}

In this paper, we introduced \tool, a structured ensemble-based approach to improve the reliability and accuracy of LLM-generated code. By leveraging pairwise similarity scoring through CodeBLEU for syntactic and semantic alignment and CrossHair-based differential analysis for behavioral consistency, our method systematically selects the most functionally correct and syntactically sound program from multiple LLM-generated candidates. Our evaluation on HumanEval and LiveCodeBench demonstrates that \tool\ consistently outperforms standalone LLMs, which highlights the potential of structured ensembling in LLM-based code generation.

In future work, we plan to extend \tool\  with more code quality metrics, related to  security, performance and maintainability, paving the way for more reliable, adaptable, and robust AI-assisted programming workflows. 

\bibliographystyle{plain}
\bibliography{Bibliography}

\end{document}